\documentclass[journal]{IEEEtran}
\usepackage{cite}

\usepackage{color,soul}
\usepackage{gensymb}
\usepackage{dsfont}

\ifCLASSINFOpdf
  \usepackage[pdftex]{graphicx}
  \graphicspath{{../pdf/}{../jpeg/}}
  \DeclareGraphicsExtensions{.pdf,.jpeg,.png}
\else
  \usepackage[dvips]{graphicx}
  \graphicspath{{../eps/}}
  \DeclareGraphicsExtensions{.eps}
\fi
\usepackage[cmex10]{amsmath}
\usepackage{amssymb}

\usepackage{euscript}

\usepackage{multirow}
\usepackage{algorithm}
\usepackage[]{algpseudocode}
\usepackage{diagbox}
\usepackage{physics}
\usepackage{tikz}

\usepackage{amsthm}

\ifCLASSOPTIONcompsoc
  \usepackage[caption=false,font=normalsize,labelfont=sf,textfont=sf]{subfig}
\else
  \usepackage[caption=false,font=footnotesize]{subfig}
\fi
\usepackage{url}

\begin{document}
%
\title{Multi-Source Data Fusion Outage Location in Distribution Systems via Probabilistic Graph Models}
%
%
%

\author{Yuxuan~Yuan,~\IEEEmembership{Graduate Student Member,~IEEE,}
	Kaveh~Dehghanpour,
	Zhaoyu~Wang,~\IEEEmembership{Senior Member,~IEEE,}
	and Fankun~Bu,~\IEEEmembership{Graduate Student Member,~IEEE}
\thanks{This work is supported by the U.S. Department of Energy Office of Electricity Delivery and Energy Reliability under DE-OE0000875.

 Y. Yuan, K. Dehghanpour, Z. Wang, and F. Bu are with the Department of
Electrical and Computer Engineering, Iowa State University, Ames,
IA 50011 USA (e-mail: yuanyx@iastate.edu; wzy@iastate.edu).
 }
}
%
%

\markboth{Submitted to IEEE for possible publication. Copyright may be transferred without notice}%
{Shell \MakeLowercase{\textit{et al.}}: Bare Demo of IEEEtran.cls for Journals}
%



\maketitle

\begin{abstract}
Efficient outage location is critical to enhancing the resilience of power distribution systems. However, accurate outage location requires combining massive \textit{evidence} received from diverse data sources, including smart meter (SM) last gasp signals, customer trouble calls, social media messages, weather data, vegetation information, and physical parameters of the network. This is a computationally complex task due to the high dimensionality of data in distribution grids. In this paper, we propose a multi-source data fusion approach to locate outage events in partially observable distribution systems using Bayesian networks (BNs). A novel aspect of the proposed approach is that it takes multi-source evidence and the complex structure of distribution systems into account using a probabilistic graphical method. Our method can radically reduce the computational complexity of outage location inference in high-dimensional spaces. The graphical structure of the proposed BN is established based on the network's topology and the causal relationship between random variables, such as the states of branches/customers and evidence. Utilizing this graphical model, accurate outage locations are obtained by leveraging a Gibbs sampling (GS) method, to infer the probabilities of de-energization for all branches. Compared with commonly-used exact inference methods that have exponential complexity in the size of the BN, GS quantifies the target conditional probability distributions in a timely manner. A case study of several real-world distribution systems is presented to validate the proposed method.

\end{abstract}

\begin{IEEEkeywords}
Approximate inference, Bayesian networks, data fusion, outage location, partially observable distribution system.
\end{IEEEkeywords}

\section{Introduction}\label{introduction}
Frequent power outages are becoming a critical issue in the U.S. In 2018, the Department of Energy estimates that outages are costing the U.S. economy $\$150$ billion annually \cite{loss}. 1.9 million customers in Midwest were affected by 1.4 million outages between August 10 and 13, 2020 \cite{midwest}. Outage detection in distribution grids is an immediate and indispensable task after service disruptions, without which utilities cannot obtain needed situational awareness for initiating repair and restoration. This suggests an urgent need of efficient approaches to shorten the time of lateral-level outage location. Traditionally, outage location inference has been done based on manual outage mapping, which in addition to the voltage and current components measured only at the substations, has mainly depended on customers' trouble calls. However, trouble calls alone are not a reliable source for outage location inference. It is estimated that only one-third of customers report the events in the first hour of outages, which might prolong the location determination process \cite{GK2014}. Also, customers might contact utilities due to temporary and individual problems other than system-level outage events, which can mislead the location process and result in additional truck rolls to verify power outages. 

One way of avoiding these problems is to rely on advanced metering infrastructure (AMI)-based techniques, which can send outage notifications at the grid-edge by leveraging the bidirectional communication function of smart meters (SMs) \cite{ZSH2018}. Researchers have dedicated great efforts to this topic. In \cite{RM2018}, a hierarchical generative model is proposed that employs SM error count measurements to detect anomalies. In \cite{ZSH2018}, a multi-label support vector machine model is developed that utilizes the state of customers' SMs to identify states of distribution lines. In \cite{SJC2015}, a two-stage method is presented to detect non-technical losses and outage events using real-time consumption data from SMs. In \cite{RA2018}, a hypothesis testing-based outage location method is developed that combines the power flow measurements and SM-based load forecasts of the nodes. In \cite{jyz2016}, by using data from SMs and fault indicators, a multiple-hypothesis method with an extended protection tree is presented to detect a fault and identify the activated protective devices. A main challenge is that most AMI-based methods require \textit{full observability} for distribution grids, i.e., SM installation for all customers. This assumption is not necessarily applicable to practical distribution systems, mostly due to utilities' budgetary limitations. To perform outage detection in partial observable systems, we have proposed a generative adversarial network (GAN)-based method to efficiently identify outage region \cite{yuanGAN}. Although this method is guaranteed to capture the maximum amount of information on outage location, it does not provide granular outage location estimation at the branch level due to the limitations of the single data source. This issue is further exacerbated considering that SM signal communication to the utilities' data centers can fail due to hardware/software malfunction and tampering \cite{RM2018}. 

Rather than using SM data, an alternative solution is to utilize other grid-independent data sources to identify outage events in real-time \cite{TS2010}. In \cite{PK2014}, weather information data is used to detect outages in overhead distribution systems employing an ensemble learning approach. In \cite{HSF2016}, a data-driven outage identification approach is proposed that extracts textural and spatial information from social media. In \cite{GA2020}, a mixed-integer linear program (MILP) is formulated to identify the topology under both outage and normal operating conditions using line flow measurements, forecasted load data, and ping measurements from a limited set of SMs. Nonetheless, the considerable uncertainty of these data sources can lead to erroneous outage location and additional costs for utilities. For example, only a part of SM last gasp signals can be delivered to the utility's data center due to hardware and software issues. Thus, to handle the limitations and uncertainties of individual data sources, this paper proposes a multi-source data fusion strategy to combine outage-related information from diverse sources for accurate outage location. 

One fundamental challenge in multi-source outage location is the computational complexity of the problem: first, note that outage location inference is the process of computing the probabilities of topology candidates after disrupting events by leveraging available information received by utilities. Estimating these probability values requires obtaining the \textit{joint probability distribution function (PDF)} of the unknown state variables and the evidence, which is a high-dimensional mathematical object. Directly quantifying this joint distribution is computationally infeasible for actual distribution systems, which requires enumerating probabilities of all possible combinations of variables. In addition, outage data sources have heterogeneous characteristics such as accuracy levels and reporting rates. Further, they may provide inconsistent and contrary information. How to integrate these data sources is a challenge.

To address these challenges and the shortcomings of the previous works in the literature, a multi-source data fusion method is presented to identify and locate the lateral-level outage events in partially observable distribution systems. To achieve this, we have adopted a \textit{probabilistic graphical modeling} approach towards data fusion to reduce the computational complexity of representing high-dimensional joint PDF of the system. The basic idea of this methodology is to first build a Bayesian network (BN) for each distribution feeder and then use this probabilistic model to infer the conditional PDF of the state of network branches and the connectivity of customer switches given the observed evidence. BNs utilize a graph-based representation as the basis for analyzing statistical relationships among random variables \cite{kaveh2016}. System topology from one-line diagram and context data, such as weather data and vegetation information, from geographic information system are used to design the architecture of the BN, as shown in Fig. 1. The graph parameters are learned empirically from historical outage data. By utilizing the proposed BN-based technique, the high-dimensional joint PDF of the system is decomposed into a set of more manageable probabilistic \textit{factors} obtained from conditional independencies. Consequently, the data fusion-based outage location process is efficiently transformed into online inference over the BN. This inference task is solved by leveraging a Gibbs sampling (GS) algorithm. As a Markov chain Monte Carlo (MCMC)-based algorithm, GS can provide a full characterization of the distribution of unknown variables by generating a sequence of samples. We have used several real-world feeder models from our utility partners to validate the performance of the proposed method. The main contributions of this paper can be summarized as follows: 
\begin{itemize}
\item The proposed method can seamlessly integrate heterogeneous data sources. Different data sources can complement each other to increase the amount of outage information, thus addressing low SM coverage or customer report rates in actual grids.
\item By exploiting the \textit{inherent conditional independencies} between the evidence and state variables in distribution systems, the exponential computational complexity of the outage location task is reduced to linear complexity in the number of variables.
\item The proposed method is robust with respect to misreports and inconsistencies in outage evidence, as the uncertainty of each data source is explicitly modeled using graph parameters. 
\item The proposed method introduces an intuitive way to incorporate the prior expert knowledge from utilities into the outage inference model.
\item The proposed method provides a principled framework to reduce \textit{overfitting} under outage data scarcity by leveraging the Bayesian hypotheses.
\end{itemize}



The rest of this paper is constructed as follows: In section \ref{motivation}, the statement of the outage location problem is described. Section \ref{pgm} presents the proposed BN-based data fusion model, along with structure selection and parameter learning schemes. An MCMC approximate inference algorithm is given in Section \ref{gb}. The numerical results are analyzed in Section \ref{result}. Section \ref{conclusion} concludes the paper with major findings.



\begin{figure}[tbp]
	\centering
	\includegraphics[width=0.9\linewidth]{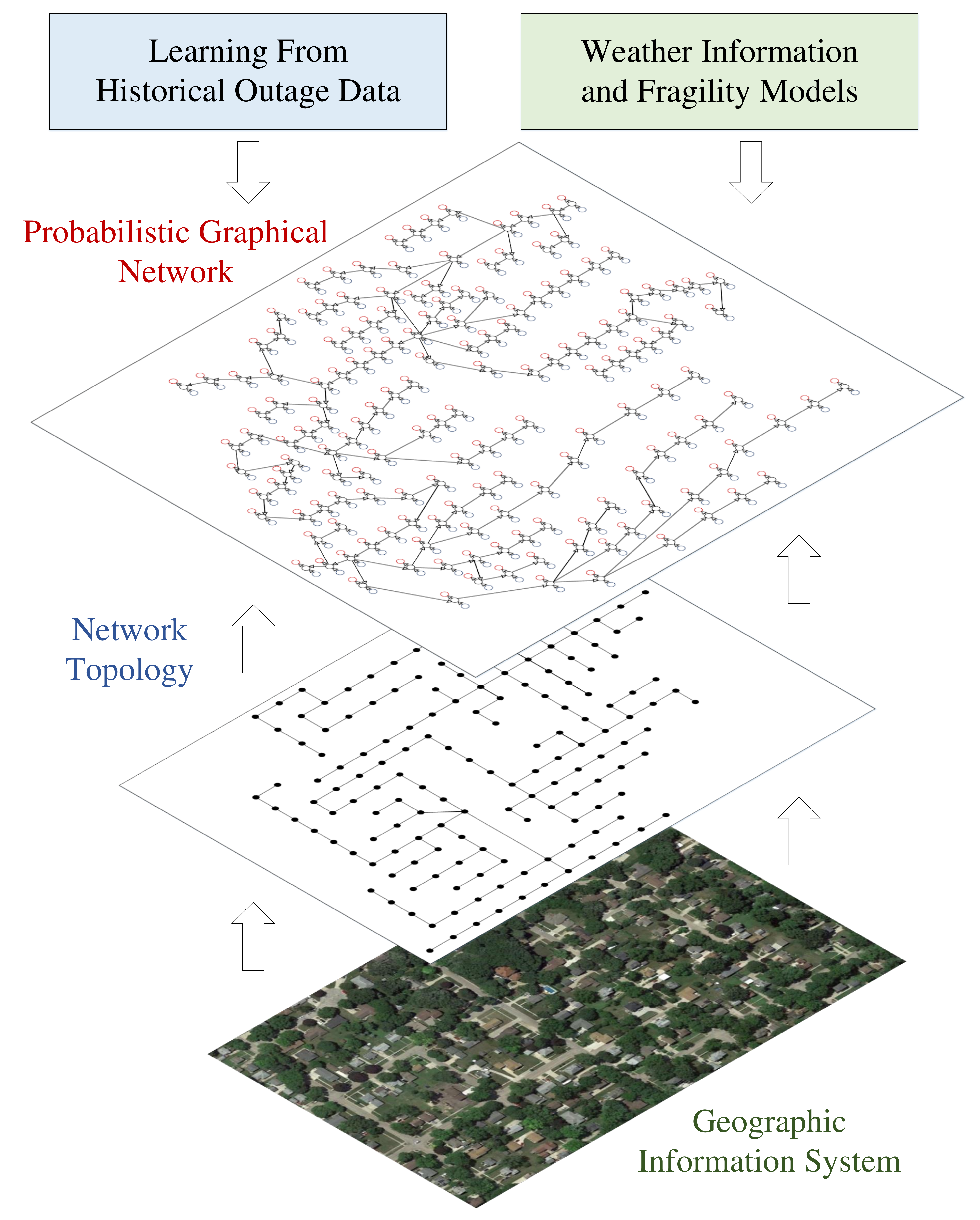}
	\caption{Graphical approach towards outage location inference.}
	\label{fig:main}
\end{figure}


\section{Outage Location Problem Statement}\label{motivation}
Considering that outage events cause topological changes in the grid, outage location is the process of inferring the probabilities of post-event operational topology candidates. The accuracy of outage location depends on the completeness of outage information. In this word, outage-related information from diverse sources are utilized for outage location, which include SM last gasp signals, customer trouble calls, social media messages, wind speed, vegetation information, and physical parameters of the grid. The rationale behind the use of wind speed and vegetation information is that 87$\%$ of major power outages are caused by extreme weather events \cite{ZSH2018}. These data sources are commonly available to utilities. Specifically, SM will automatically send the last gasp signal to the head end system of the AMI after the power fails. Trouble calls and social media messages are reported by customer's phones and Twitter. Wind speed, vegetation information, and the physical parameters of the grid can be found from the neighboring land-based station, the geographic information system, and the grid model, respectively. Thus, our method does not require to install extra metering devices, such as SMs, line flow measurements, and micro-phasor measurement units ($\mu$PMUs), for data collection. This ensures the practicability of the proposed method for actual distribution systems. Further, since the proposed method is scalable and adaptive, utilities are able to add or remove data sources according to their situations. For example, in the absence of extreme weather events, utilities can remove vegetation information and wind speed to reduce the complexity of the model, as these two data sources may not have a significant impact on outage detection and location during normal weather. After data collection, last gasp signals, customer trouble calls, wind speed, vegetation information, and physical parameters can be directly transformed into outage evidence as input to the proposed model. For social media messages, a natural language processing tool is required to extract outage-related words, as proposed in our previous work \cite{HSF2016}. Then, social media messages are converted into binary outage evidence, similar to customer trouble calls and last gasp signals.

Given the multi-source evidence, $\pmb{E}$, this inference process is mathematically formulated using the \textit{Bayes estimator} \cite{ds2018}, where the conditional PDF of network topology, $Y$, given the set of evidence is represented as $P(Y|\pmb{E})$ and calculated in terms of the joint distribution of $Y$ and $\pmb{E}$, denoted by $P(Y,\pmb{E})$. Note that in this paper, random vectors/matrices are represented with bold letters. The most probable candidate topology, which also determines the location of the outage event, is obtained by maximizing this conditional PDF, as:
\begin{equation}
\label{eq:be}
Y^*=\operatorname*{argmax}_Y P(Y|\pmb{E})=\frac{P(Y,\pmb{E})}{\sum_{y}P(y,\pmb{E})}
\end{equation}
where, $Y^*$ is the most likely network topology after the outage. $Y$ is a multinomial variable which is represented in terms of the states of primary network branches ($\pmb{D}$) and the connection of customer switches ($\pmb{C}$), as $Y = \{\pmb{D}, \pmb{C}\}$. Here, $\pmb{D}=[D_1,...,D_k]$, where $k$ is the number of branches in the feeder and $D_i$ is a binary variable representing the connectivity state for the $i$'th branch in the feeder: $D_i=0$ means that the branch is \textit{energized}. In other words, there is an uninterrupted path between the branch and the substation. $D_i=1$ indicates that the branch is \textit{de-energized}. Similarly, $\pmb{C}=[\pmb{C_1},...,\pmb{C_k}]$, with $\pmb{C_i}$ representing the set of connection states for all the customers that are supplied by the $i$'th branch. Hence, $\pmb{C_i}=[C_i^1,...,C_i^{z_i}]$, where $z_i$ is the total number of customers that are connected to the $i$'th branch, and $C_i^j$ is the state of the $j$'th customer: $C_i^j=0$ means that the customer is energized, and $C_i^j=1$ implies that the customer is de-energized. Note that the pre-outage topology is determined by assigning $0$ to all the state variables (i.e., all branches are energized and customers are energized). Thus, $P(Y|\pmb{E})$ in \eqref{eq:be} can be rewritten in terms of the joint PDF of the newly-defined variables, $P(\pmb{D},\pmb{C},\pmb{E})$, as follows \cite{BN2013}:
\begin{equation}
\label{eq:be0}
P(Y|\pmb{E})=P(\pmb{D},\pmb{C}|\pmb{E})=\frac{P(\pmb{D},\pmb{C},\pmb{E})}{\sum_{\pmb{d}}\sum_{\pmb{c}}P(\pmb{d},\pmb{c},\pmb{E})}.
\end{equation}
Using \eqref{eq:be0}, the maximization over topology candidates can be conveniently transformed into finding the best values for the individual branch/customer states belonging to $\{\pmb{D},\pmb{C}\}$ using their conditional PDFs, $P(D_i|\pmb{e})$ and $P(C_i^j|\pmb{e})$. These conditional PDFs are obtained $\forall i,j$ using a marginalization process over the joint PDF, as follows \cite{KD2009}:
\begin{equation}
\label{eq:be1-1}
P(D_i|\pmb{E})=\sum_{\{\pmb{d},\pmb{c}\}\setminus d_{i}}P(\pmb{D},\pmb{C}|\pmb{E})= \sum_{\{\pmb{d},\pmb{c}\}\setminus d_{i}}\frac{P(\pmb{D},\pmb{C},\pmb{E})}{\sum_{\pmb{d}}\sum_{\pmb{c}}P(\pmb{d},\pmb{c},\pmb{E})}
\end{equation}
\begin{equation}
\label{eq:be1-2}
P(C_i^j|\pmb{E})=\sum_{\{\pmb{d},\pmb{c}\}\setminus c_{i}^j}P(\pmb{D},\pmb{C}|\pmb{E})= \sum_{\{\pmb{d},\pmb{c}\}\setminus c_{i}^j}\frac{P(\pmb{D},\pmb{C},\pmb{E})}{\sum_{\pmb{d}}\sum_{\pmb{c}}P(\pmb{d},\pmb{c},\pmb{E})}.
\end{equation}

Solving \eqref{eq:be1-1}-\eqref{eq:be1-2} requires quantifying the joint PDF, $P(\pmb{D},\pmb{C},\pmb{E})$. Considering the complexity of distribution grids, obtaining the explicit representation of this joint PDF is unmanageable for two reasons: (I) a complete description of $P(\pmb{D},\pmb{C},\pmb{E})$ induces an exponential complexity in the order of $2^r-1$, where $r$ is the total cardinality of all the unknown variables, $r = |\pmb{D}| + |\pmb{C}|$. Hence, modeling this joint PDF using brute-force search over all possible combinations of branch/customer states is computationally infeasible for large-scale distribution systems. (II) Due to the outage data scarcity in distribution grids, it is impossible to acquire enough historical data to robustly estimate the massive number of parameters of this joint distribution. One solution is to use \textit{naive classification} by assuming full independence among all evidence and unknown state variables \cite{KD2009}. However, this assumption is not applicable to practical distribution systems and may lead to severe misclassification due to overfitting.

\section{BN-Based Data Fusion Model}\label{pgm}

To counter computational complexity and overfitting in the outage location inference, we propose a BN-based method to provide a compact representation of the high-dimensional joint PDF, $P(\pmb{D},\pmb{C},\pmb{E})$. To do this, our method exploits conditional independencies between random variables, $\{\pmb{D},\pmb{C},\pmb{E}\}$, to decompose the joint PDF into a set of \textit{factors} with significantly smaller size. Using this computationally efficient BN-based approach, we can infer the conditional PDF of the state of each primary branch and the customer switch given outage-related evidence from various data sources, shown in \eqref{eq:be1-1}-\eqref{eq:be1-2}, to rapidly identify the location of lateral-level outage events.

\subsection{Factorization of the Joint PDF and BN Representation}
The main idea of a BN-based representation is to use conditional independencies, encoded in a graph structure, to compactly break down high-dimensional joint PDFs with a set of factors. Here, a factor refers to a low-dimensional and more manageable conditional PDF that is determined by two components: a \textit{child} variable, such as $D_i$ and a number of \textit{parent} variables denoted by $Pa(\cdot)$, such as $Pa(D_i)$. Parents represent the direct causal sources of influence for a child variable. In other words, each child is a stochastic function of its parents \cite{KD2009}. Thus, if the values of the parents are known, then the child variable becomes conditionally independent of random variables that do not directly influence it in a causal manner. It can be shown that by using chain rule over these conditional independencies, defined by parent-child relationships, the joint PDF of a set of random variables can be simplified as the multiplication of the identified factors \cite{KD2009}. In the outage location problem, this factorization leads to the following data fusion representation for the joint PDF:
\begin{equation}
\begin{split}
\label{eq:cr3}
&P(\pmb{D},\pmb{C},\pmb{E}) = \prod_{i=1}^{k}P(D_i|Pa(D_i))\prod_{i=1}^{k}\prod_{j=1}^{z_i}P(C_i^j|Pa(C_i^j))\\
&\times\prod_{i=1}^{u}P(E_{i,j}^h|Pa(E_{i,j}^h))\prod_{i=1}^{u}P(E_{i,j}^m|Pa(E_{i,j}^m))
\end{split}
\end{equation}
where, $u = |\pmb{E}|$, and the factors are: $P(D_i|Pa(D_i))$, $P(C_i^j|Pa(C_i^j))$, $P(E_{i,j}^h|Pa(E_{i,j}^h))$, and $P(E_{i,j}^m|Pa(E_{i,j}^m))$, $\forall i,j$.  $E_{i,j}^h$ denotes the human-based evidence from the customer-side, including trouble calls and social media messages; $E_{i,j}^m$ represents meter-based evidence from customer-side, such as smart meter last gasp signals. Compared with the original model in \eqref{eq:be0} that requires $2^r-1$ independent parameters, the new representation in \eqref{eq:cr3} only needs $\sum_{i=1}^{k}2^{|Pa(D_i)|} + \sum_{i=1}^{k}\sum_{j=1}^{z_i}2^{|Pa(C_i^j)|} + \sum_{i=1}^{n}2^{|Pa(E_{i,j}^h)|} + \sum_{i=1}^{n}2^{|Pa(E_{i,j}^m)|}$ parameters. It can be observed that the number of parameters in the new representation is a function of size of parents for each variable. Considering that the number of variables' parents is typically small, the new representation achieves a radical complexity reduction in outage location inference.


As a directed acyclic graph, BN offers a convenient way of representing the factorization \eqref{eq:cr3}. Accordingly, the random variables, $\{\pmb{D},\pmb{C},\pmb{E}\}$, are represented as the \textit{vertices} of the BN. Using the identified factors in \eqref{eq:cr3}, the vertices of the BN are connected by drawing \textit{directed edges} that start from parent vertices and end in child vertices. BN provides a graphical way for encoding conditional independencies defined by the factors as follows: any vertex, $X$, is conditionally independent of its \textit{non-descendant vertices} in the graph, $Nd(X)$, if the values of its parents are known. This is symbolically denoted by $(X \perp Nd(X)|Pa(X))$ \cite{NB2017}. $Nd(X)$ is the set of the vertices of the BN, excluding parents of $X$, to which no directed path exists originating from $X$. $A \perp B$ means that $A$ and $B$ are marginally independent.

\subsection{BN Structure Development and Parameterization} \label{BNSP}
Developing a BN requires discovering the structure of the graph and the parameters of the conditional PDFs. To do this, a knowledge discovery-based method is utilized in this paper. The basic idea of this method is to reveal the conditional independencies between variables using the grid topology information and causal relationships. Then, the conditional PDFs (i.e., factors) are parameterized based on the available statistical outage information from the literature \cite{HSF2016,TS2010,ppt2014}. For concreteness, the parent-child variables of each factor in \eqref{eq:cr3} are described as follows:

\begin{figure}[tbp]
	\centering
	\includegraphics[width=3.5in]{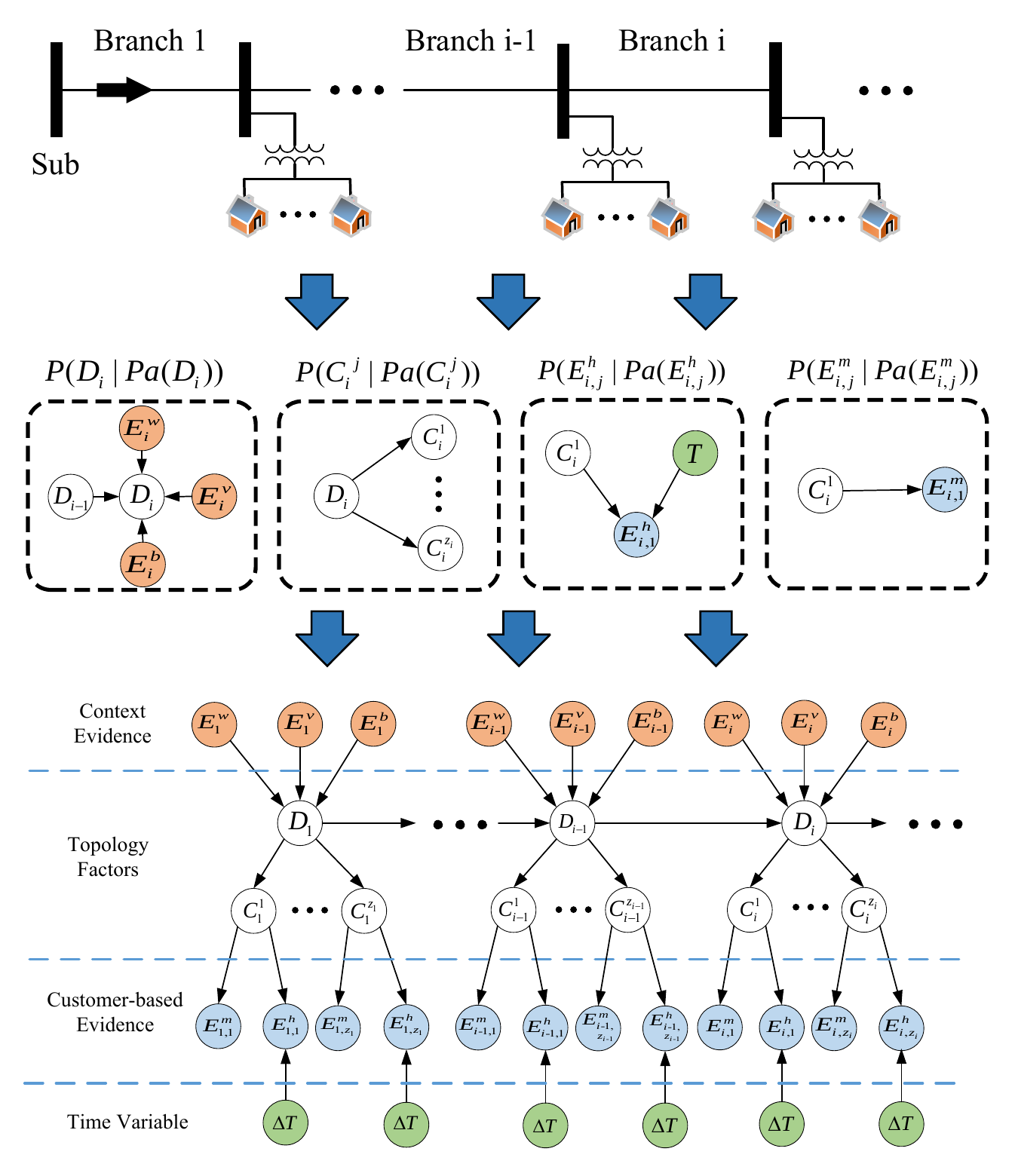}
	\caption{BN of a typical radial distribution system.}
	\label{fig:pbn}
\end{figure}

(1) Factor $P(D_i|Pa(D_i))$ represents the conditional independencies of the form $D_{i} \perp Nd(D_{i})|Pa(D_{i})$. The parents of branch state variable are selected as $Pa(D_{i}) = \{D_{i-1},E^w_i,E^v_i,E^b_i\}$, as shown in Fig. \ref{fig:pbn}. Here, $D_{i-1}$ is the state of the neighboring upper-stream branch. $\{E^w_i,E^v_i,E^b_i\}$ are the evidence for the $i$'th branch, where $E^w_i$ denotes wind speed; $E^v_i$ is vegetation information, which contains vegetation species-specific constants and tree diameter; $E^b_i$ represents the $i$'th branch's physical parameters including the length of conductors and the number of poles. Based on this parent selection scheme for branch state variables, $Nd(D_{i})$ includes all the variables that are not downstream of the $i$'th branch in the feeder (see Fig. \ref{fig:pbn}). To show the direct causal influences of these four variables on $D_i$, two cases are described: $D_{i-1}=1$ and $D_{i-1}=0$. 

In the first case, when the parent branch is de-energized, then $D_{i}=1$ with \textit{probability 1}. Consequently, all the variables in the path from the substation to $D_{i-1}$, represented with $\{D_1,...,D_{i-2}\}$, are conditionally independent from $\{D_i\}$ given $D_{i-1}=1$. The intuition behind this is that in radial networks there can only be a unique path between the substation and each branch; if this path is interrupted at any arbitrary point in $\{D_1,...,D_{i-2}\}$, we can automatically conclude $D_{i-1}=1$ regardless of the location of outage in the path. Hence, considering the binary nature of variable $D_i$, the conditional PDF, $P(D_i|D_{i-1}=1,E_{i}^w,E_{i}^v,E_{i}^b)$, can be formulated as:
\begin{equation}
\label{eq:d_cdf_1}
\begin{split}
&P(D_i=1|D_{i-1}=1,E_{i}^w,E_{i}^v,E_{i}^b) = 1\\
&P(D_i=0|D_{i-1}=1,E_{i}^w,E_{i}^v,E_{i}^b) = 0.
\end{split}
\end{equation}

In the second case, if the neighboring upper-stream branch is energized, then all upstream branches of the $i$'th branch are also energized with probability 1, and have not been impacted by outage, $\{D_1=0,...,D_{i-2}=0\}$. In this case, $D_i=1$ will only occur when this branch is damaged. As demonstrated concretely in \cite{sma2018}, the majority of branch damage is caused by tree contacts to power lines and broken poles due to high wind speed. Thus, three context variables $E^w_i$, $E^v_i$ and $E^b_i$ are serve as causal evidence for the $i$'th branch state to estimate the probability of outage at the $i$'th branch. The conditional PDF, $P(D_i|D_{i-1}=0,E_{i}^w,E_{i}^v,E^b_i)$, can be formulated as a Bernoulli distribution as follows:
\begin{equation}
\label{eq:d_cdf_2}
P(D_i|D_{i-1}=0,E_{i}^w,E_{i}^v,E^b_i) =
\begin{cases}
P_l^i &\mathrm{for} \ D_i = 1\\
1-P_l^i &\mathrm{for} \ D_i = 0
\end{cases} 
\end{equation}
where, the probability of failure for branch $i$, denoted as $P_l^i$, is a function of $E^w_i$, $E^v_i$, and $E^b_i$. To formulate this function, a fragility model is leveraged \cite{sma2018}. Basically, the fragility model is a series model with the fragility analysis of each pole and conductor within the branch:
\begin{equation}
\begin{split}
\label{eq:frag_model}
P_l^i=1-\prod^L_{d=1}(1-\Phi(\frac{\ln{(\frac{E^w_i}{\chi})}}{\xi}))\prod^K_{f=1}(1-P_f(E^w_i,E^v_i))
\end{split}
\end{equation}
where, $L$ is the number of distribution poles used for supporting branch $i$, $K$ is the number of conductor wires between two neighboring poles at the $i$'th branch, $\Phi$ is the standard normal probability integral, $\chi$ is the median of the fragility function, $\xi$ is the logarithmic standard deviation of intensity measure, and $P_f(E^w_i,E^v_i)$ represents the failure probability for conductor $f$ of branch $i$ which is modeled as follows:  
\begin{equation}
\label{eq:frag_model_2}
P_f=(1-p_u)\max{\{\min\{{\frac{F_{wind,f}(E^w_i)}{F_{no,f}(E^w_i)}, 1\}},\alpha\cdot{P_f(E^v_i)}\}}
\end{equation}
where, $p_u$ is the probability of conductor $f$ being underground, $F_{wind,f}(E^w_i)$ represents the wind force loading on the conductor and $F_{no,f}(E^w_i)$ demonstrates the maximum perpendicular force of the conductor wire determined as shown in \cite{AM2015}. $\alpha$ describes the average tree-induced damage probability of overhead conductor, and $P_f(E^v_i)$ is the fallen tree-induced failure probability of conductor $f$ computed as in \cite{sma2018}.
Hence, for the case $D_{i-1}=0$, equations \eqref{eq:frag_model} and \eqref{eq:frag_model_2} are utilized to estimate the probability of outage for branch $i$ given the context variables $E^w_i$, $E^v_i$, and $E^b_i$. To summarize, the conditional PDFs given in equations \eqref{eq:d_cdf_1} and \eqref{eq:d_cdf_2} fully determine the factors of the form $P(D_i|Pa(D_i))$.

(2) Factor $P(C_i^j|Pa(C_i^j))$ represents the conditional PDF of the status of customer $j$ given parent variables. The parent of customer state variable is selected as $Pa(C_i^j) = \{D_{i}\}$ (see Fig. \ref{fig:pbn}). Here, $D_i$ is the state of the immediate upper-stream branch that supplies the $j$'th customer. To show the casual relationship between $C_i^j$ and $D_i$, two cases are considered: $D_i=1$ and $D_i=0$. 

In the first case, if the primary branch is de-energized, the probability of $C_i^j=1$ is $1$ due to the radial structure of the feeder. Utilizing this deterministic relationship, $P(C_i^j|D_i=1)$ can be written as follows:
\begin{equation}
\begin{split}
\label{eq:ed}
&P(C_i^j=1|D_{i}=1) = 1\\
&P(C_i^j=0|D_{i}=1) = 0.
\end{split}
\end{equation}

In the second case, if the primary branch is energized, then the path between the substation and the $i$'th branch is active. In this case, customer outage, $C_i^j=1$, can only be caused by overloading/faults at the customer-side occurring with probability $\pi_2$. This case is represented using a Bernoulli distribution adopted from statistical outage information \cite{Report}:
\begin{equation}
\begin{split}
\label{eq:ed}
&P(C_i^j|D_i=0) =
\begin{cases}
\pi_2 &\mathrm{for} \ C_i^j = 1\\
1-\pi_2 &\mathrm{for} \ C_i^j = 0.
\end{cases} 
\end{split}
\end{equation}
To account for the uncertainty of parameter $\pi_2$, a beta distribution is defined with user-defined hyper-parameters $\alpha_2$ and $\beta_2$:
\begin{equation}
\label{eq:ed1}
\pi_2\sim Beta(\alpha_2,\beta_2)=\gamma_2\pi_2^{\alpha_2-1}(1-\pi_2)^{\beta_2-1} \\
\end{equation}
where, $\gamma_2$ is a normalizing constant and defined as $\gamma_2=\Gamma(\alpha_2+\beta_2)$ with $\Gamma = \int_0^{\infty}t^{x-1}e^{-t}dt$ \cite{KD2009}. 


(3) Factor $P(E_{i,j}^h|Pa(E_{i,j}^h))$ represents the conditional independencies $E_{i,j}^h \perp Nd(E_{i,j}^h)|Pa(E_{i,j}^h)$. The parents of human-based evidence, $E_{i,j}^h$, are selected as $Pa(E_{i,j}^h) = \{C_i^j, \Delta{T}\}$, as shown in Fig. \ref{fig:pbn}. $\Delta{T}$ refers to the time elapsed after the outage occurrence. More precisely, $\Delta{T}$ embodies the time period that utilities need to wait before outage reports are issued \cite{JY2020}. This is to avoid false alarms from customers due to temporary events. It is clear that there is a trade-off between the amount of human-based evidence and waiting time of outage location inference. For example, when the observability of the actual feeder is extremely low, the utility may increase $\Delta T$ to receive more human-based evidence for outage location inference. Within the $\Delta T$ period, the time at which the human-based evidence is received, $t$, after outage occurrence at time, $T_0$, is distributed according to an exponential distribution as shown in \cite{TS2010}:
\begin{equation}
\label{eq:ex1}
f(t|C_i^j=1)=\lambda_1 e^{-\lambda_1 (t-T_0)}.
\end{equation}
Thus, given $\Delta{T}$, the probability of $P(E_{i,j}^h|C_i^j=1, (t-T_0)\leq \Delta{T})$ can be calculated as:
\begin{equation}
\begin{split}
\label{eq:ex2}
&P(E_{i,j}^h=1|C_i^j=1,t-T_0 \leq \Delta{T}) \\
&=\int_{0}^{\Delta{T}} \lambda_1 e^{-\lambda_1 t'}\mathrm{d}t'=-e^{-\lambda_1 \Delta{T}}+1.
\end{split}
\end{equation}
Hence, the factor $P(e_{i,j}^h|C_i^j,t\leq \Delta{T})$ is obtained as follows:
\begin{equation}
\begin{split}
\label{eq:he}
&P(E^h_{i,j}|C_i^j,t \leq \Delta{T})=\\
&\begin{cases}
-e^{-\lambda_1 \Delta{T}}+1 &\mathrm{for} \ E_{i,j}^h = 1,C_i^j=1\\
e^{-\lambda_1 \Delta{T}} &\mathrm{for} \ E_{i,j}^h = 0,C_i^j=1\\
\pi_3  &\mathrm{for} \ E_{i,j}^h = 1,C_i^j=0\\
1-\pi_3 &\mathrm{for} \ E_{i,j}^h = 0,C_i^j=0\\
\end{cases} 
\end{split}
\end{equation}
where, $\pi_3$ denotes a small user-defined value to take into account the possibility of false positives, such as illegitimate trouble call and social media data processing errors. 

(4) Factor $P(E_{i,j}^m|Pa(E_{i,j}^m))$ is the conditional independencies $E_{i,j}^m \perp Nd(E_{i,j}^m)|Pa(E_{i,j}^m)$. Compared to the human-based signals $E_{i,j}^h$, AMI-based notification mechanism will be delivered almost instantaneously to the utilities. Thus, the parent of meter-based evidence is selected as $Pa(E_{i,j}^m) = \{C_i^j\}$ (see Fig. \ref{fig:pbn}). When the state of customer switch is known, $E_{i,j}^m$ becomes conditionally independent of the remaining variables, as encoded by the factor:
\begin{equation}
\begin{split}
\label{eq:me}
P(E^m_{i,j}|C_i^j)=
\begin{cases}
\pi_4 &\mathrm{for} \ E_{i,j}^m = 1,C_i^j=1\\
1-\pi_4 &\mathrm{for} \ E_{i,j}^m = 0,C_i^j=1\\
\pi_5  &\mathrm{for} \ E_{i,j}^m = 1,C_i^j=0\\
1-\pi_5 &\mathrm{for} \ E_{i,j}^m = 0,C_i^j=0\\
\end{cases} 
\end{split}
\end{equation}
where, $\pi_4$ and $\pi_5$ represent the AMI communication reliability and the SM malfunction probability values, respectively. For concreteness, $\pi_4$ is the probability that the last gasp can be delivered to the utilities correctly for outage notification. $\pi_5$ is  the probability that the SM loses power due to its own failure and sends a last gasp signal. In this work, the values of these two parameters are determined based on the historical outage report. Considering the size of the historical data is limited, beta distributions are used to model the uncertainty of these two parameters as follows:
\begin{equation}
\begin{split}
\label{eq:ed1}
\pi_4\sim Beta(\alpha_4,\beta_4)=\gamma_4\pi_4^{\alpha_4-1}(1-\pi_4)^{\beta_4-1} \\
\pi_5\sim Beta(\alpha_5,\beta_5)=\gamma_5\pi_5^{\alpha_5-1}(1-\pi_5)^{\beta_5-1}.
\end{split}
\end{equation}

\begin{algorithm}
\caption{Outage Location Inference using GS}\label{alg:gbl}
\begin{algorithmic}[1]
\Require: {BN $G$; iteration number $M$; evidence $\pmb{E}$;}
\State {Randomly generate i.i.d. samples $\pmb{x}^{(0)}\leftarrow\{D_i=d_i^{(0)},...,C_i^j=c_i^{j,(0)}, \forall i,j\}$ from uniform distribution; $\pmb{x}^{(0)}\leftarrow\pmb{x}^{(0)}\cup \pmb{E}$}
    \For {$\tau = 0,...,M$}
        \For {$i = 1,...,|\pmb{D} + \pmb{C}|$}
            \State {Select one random variable $X_i \in \{\pmb{D},\pmb{C}\}$}
            \State {$\pmb{x_{-i}}^{(\tau)}\leftarrow \pmb{x}^{(\tau)} - x_i^{(\tau)}$}
            \State {Obtain $Pa(X_i)$ and $Ch(X_i)$ from $G$}
            \State {$P_\Phi(X_i|\pmb{x_{-i}}^{(\tau)})$$\leftarrow$$\frac{P(X_i|Pa(X_i))P(Ch(X_i)|X_i)}{\sum_{x_i}P(X_i|Pa(X_i))P(Ch(X_i)|X_i)}$}
            \State {Draw a new sample, $x_i^{(\tau+1)}$ $\sim$ $P_\Phi(X_i|x_{-i})$}
            \State {$x_i^{(\tau+1)} \gets x_i^{(\tau)}$}
        \EndFor    
    \EndFor
    \State {Return sample vectors: $\pmb{d_i}=\{d_i^{(0)},...,d_i^{(M)}\}$ and $\pmb{c_i^{j}}=\{c_i^{{j},(0)},...,c_i^{{j},(M)}\}$, $\forall i,j$}
    \State {$P(D_i=1|\pmb{E})\leftarrow \frac{\sum_{\tau=0}^M{d_i^{(\tau)}}}{M}, \forall i$} 
    \State {$P(C_i^j=1|\pmb{E})\leftarrow \frac{\sum_{\tau=0}^M{c_i^{j,(\tau)}}}{M}, \forall i,j $}
    \State {If $P(D_i=1|\pmb{E})\leq 0.5 \Longrightarrow D_i=1, \forall i$; if $P(C_i^j=1|\pmb{E})\leq 0.5 \Longrightarrow C_i^j=1, \forall i,j$}
    \State {Select the nearest de-energized branch as the outage location}
\end{algorithmic}
\end{algorithm}

\section{BN-Based Outage Location Inference Using GS}\label{gb}
After construction and parameterization of the BN, the data fusion outage location process is efficiently transformed into a probabilistic inference over the graphical model. However, even if $P(\pmb{D},\pmb{C},\pmb{E})$ is simplified, solving \eqref{eq:be1-1}-\eqref{eq:be1-2} still requires calculating computationally expensive summation operations $\sum_{\pmb{d}}\sum_{\pmb{c}}P(\pmb{d},\pmb{c},\pmb{E})$ over all nodes of the graph simultaneously, which is not scalable for large-scale distribution grids \cite{KD2009}. To address this, a GS algorithm is used to perform the inference task over the BN \cite{ds2019}.

\subsection{GS Algorithm}\label{method}

GS is an MCMC-based approximate inference method, which allows one to provide a good representation of a PDF by leveraging random variable instantiations, without knowing all the distribution's mathematical properties \cite{ds2019}. The key advantage of this method is that it employs univariate conditional distributions for sampling, which eliminates the dependency on the dimension of the random variable space. Thus, compared to the commonly-used exact inference methods, such as variable elimination and clique trees, GS is insensitive to the size of BN \cite{BN2013}. This indicates that the GS method is especially beneficial for complex real-world applications. 

When an outage occurs, the de-energization probabilities of branches/customers are inferred using the GS algorithm and the BN structure. To do this, first, all the outage evidence from the customer-side,
$\{E^h_{1,1},...,E^h_{z_k,k},E^m_{1,1},...,E^m_{z_k,k}\}$, is collected after $\Delta{T}$ has elapsed: if utilities receive trouble call/tweet or last gasp signal from the $j$'th customer at branch $i$, the corresponding evidence $E^h_{i,j}$ or $E^m_{i,j}$ is set to $1$. The branch-level evidence, $\{E^w_1,...,E^w_k,E^v_1,...,E^v_k,E^b_1,...,E^b_k\}$, is obtained from the utilities' data centers and weather information systems. After collecting all evidence, arbitrary initial samples are randomly assigned to all the unknown state variables $\{\pmb{D},\pmb{C}\}$: $[D_1=d_1^{(0)},...,D_k=d_k^{(0)},C_1^1=c_1^{1,(0)},...,C_k^{z_k,(0)}]$. Then, an arbitrary state variable is selected as the sampling starting point, e.g., $D_i$. At iteration $\tau+1$ of GS, following the structure of the BN, the assigned samples to the parents and children of $D_i$ are inserted into a local Bayesian estimator \cite{BN2013}, as shown in \eqref{eq:gs1}, to approximate the conditional PDF of $D_i$ given the latest samples:
\begin{equation}
\label{eq:gs1}
P_\Phi(D_i|\pmb{d_{-i}}^{(\tau)})=\frac{P(D_i|Pa(D_i))P(Ch(D_i)|Pa(Ch(D_i)))}{\sum\limits_{D_i}P(D_i|Pa(D_i))P(Ch(D_i)|Pa(Ch(D_i)))}
\end{equation}
where, $\pmb{d_{-i}}^{(\tau)}$ is all the latest samples except for $d_i$, including values of evidence variables, and:
\begin{equation}
\label{eq:f1}
P(D_i|Pa(D_i) = P(D_i|d_{i-1}^{(\tau)},E^w_i,E^v_i,E^b_i)
\end{equation}
\begin{equation}
\begin{split}
\label{eq:gs1}
&P(Ch(D_i)|Pa(Ch(D_i)))=\\
&P(d_{i+1}^{(\tau)}|D_i,E^w_i,E^v_i,E^b_i)\prod_{j=1}^{z_i}P(c_i^{j,(\tau)}|D_i).
\end{split}
\end{equation}
Hence, $P_\Phi(D_i|\pmb{d_{-i}}^{(\tau)})$ can be directly calculated using the determined factors, \eqref{eq:d_cdf_1}-\eqref{eq:ed1}, in Section \ref{BNSP}. Note that because $P_\Phi(D_i|\pmb{d_{-i}}^{(\tau)})$ is a PDF over a single random variable given the samples assigned to all the others, this computation can be performed efficiently. Utilizing $P_\Phi(D_i|\pmb{d_{-i}}^{(\tau)})$, a new sample $D_i\leftarrow d_i^{(\tau+1)}$ is drawn using the inverse transform method \cite{KD2009} to replace $d_i^{(\tau)}$. Then, the algorithm moves to a next non-evidence variable of BN to perform the local sampling process (see \eqref{eq:gs1}). When all the unknown variables of the BN have been sampled once, one iteration of GS is complete. This process is able to propagate the information across the BN and combine the data from diverse sources to infer the location of outage efficiently. The sampling process is repeatedly applied until a sufficient number of random samples are generated for the unknown variables, $\{\pmb{D},\pmb{C}\}$. It has been theoretically proved that the approximate PDFs, $P_\Phi(\cdot)$, are guaranteed to approach the target conditional PDFs, $P(D_i|\pmb{E})$ and $P(C_i^j|\pmb{E})$, defined in \eqref{eq:be1-1}-\eqref{eq:be1-2} \cite{KD2009}. Thus, $P(D_i|\pmb{E})$ and $P(C_i^j|\pmb{E})$ can be estimated by counting the samples generated by the GS algorithm. As an example, $P(D_i = 1|\pmb{E})$ is estimated as follows: 
\begin{equation}
\label{eq:gs_f}
P(D_i=1|\pmb{E}) \approx \frac{\sum_{\tau=0}^{M}{d_i^{\tau}}}{M}
\end{equation}
where, $M$ is the number of iterations. After the GS process, the most likely value of each branch/customer state is determined based on the obtained approximated conditional PDFs to solve \eqref{eq:be}. To achieve this, due to the binary nature of the state variables, a $0.5$ threshold is used, e.g. $P(D_i=1|\pmb{E}) \leq 0.5$ indicates branch $i$ is energized. After the connectivity states of all the branches/customers are inferred, the location of outage events are obtained by selecting the nearest de-energized branch to the substation. See Algorithm \ref{alg:gbl} for details.

\begin{table*}
\centering
\setlength{\tabcolsep}{1.6mm}
\renewcommand\arraystretch{1.6}
\caption{Outage Location Observability Sensitivity Analysis}
\begin{tabular}{ccccccc}
\hline\hline
System Name & Observability & Branch-level Accuracy & Branch-level Precision & Branch-level Recall & Branch-level $F_1$ & System-level Accuracy\\
\hline
\multirow{3}*{51-Node Test Feeder} & {\centering} 25$\%$ & 99.05$\%$ & 86.48$\%$ & 99.56$\%$ & 90.65$\%$ & 69.73$\%$\\
~ & 50$\%$ & 99.65$\%$ & 92.77$\%$ & 99.82$\%$ & 95.07$\%$ & 83.93$\%$\\
~ & 75$\%$ & 99.89$\%$ & 98.38$\%$ & 100$\%$ & 98.93$\%$ & 96.33$\%$\\
\hline
\multirow{3}*{77-Node Test Feeder} & 25$\%$ & 98.7$\%$ & 83.47$\%$ & 98.88$\%$ & 88.05$\%$ & 69.5$\%$\\
~ & 50$\%$ & 99.41$\%$ & 92.43$\%$ & 98.86$\%$ & 94.32$\%$ & 86.6$\%$\\
~ & 75$\%$ & 99.60$\%$ & 92.82$\%$ & 99.89$\%$ & 95.24$\%$ & 88.1$\%$\\
\hline
\multirow{3}*{106-Node Test Feeder} & 25$\%$ & 98.92$\%$ & 83.91$\%$ & 99.05$\%$ & 88.61$\%$ & 69.6$\%$\\
~ & 50$\%$ & 99.58$\%$ & 91.11$\%$ & 99.54$\%$ & 94.1$\%$ & 80.9$\%$\\
~ & 75$\%$ & 99.92$\%$ & 98.19$\%$ & 100$\%$ & 98.88$\%$ & 92.6$\%$\\
\hline\hline
\end{tabular}
\label{table:1.1}
\end{table*}

\subsection{GS Calibration Process}\label{cd_gs}
One challenge in GS is how to determine the number of iterations, $M$. In general, if the iterations have not proceeded long enough, the sampling may grossly misrepresent the target distributions, thus decreasing the inference accuracy. In contrast, if the value of M is large enough, the theory of MCMC guarantees that the stationary distribution of the samples generated using the GS algorithm \cite{BN2013}. However, such a strategy leads to high computational time, which increases outage duration and cost. Hence, by using GS, a trade-off exists between the accuracy and computational time of outage location. To find a reasonable maximum iteration number for a specific BN, a \textit{potential scale reduction} factor, $R$, is utilized to diagnose the convergence of the GS at different numbers of iterations \cite{BU2017}. The basic idea is to measure between- and within-sequence variances of generated sample sequences. Specifically, for each M, we start with $n$ sample sequences produced by the GS for each unknown variable in the BN. After discarding the samples generated in the warm-up period, each sequence is divided into two halves of the same size, $m$, and used to complement the original sequences. All the sample sequences are concatenated into a matrix of size $2n\times m$, denoted as $\pmb{\theta}$. Utilizing this matrix, the between-sequence and within-sequence variances are calculated as follows:
\begin{equation}
\label{eq:index1}
B_i=\frac{m}{2n-1}\sum_{j=1}^{2n}(\bar{\pmb{\theta}}_{.j}-\bar{\pmb{\theta}}_{..})^2
\end{equation}
\begin{equation}
\label{eq:index2}
V_i = \frac{1}{2n}\sum_{j=1}^{2n}s_j^2
\end{equation}
where, $B_i$ is the between-sequence variance of variable $i$, $V_i$ is the within-sequence variance of variable $i$, $\bar{\pmb{\theta}}_{.j}$ is the within-sequence means that can be calculated using $\bar{\pmb{\theta}}_{.j}=\frac{1}{m}\sum_{i=1}^{m}\pmb{\theta}_{ij}$. $\bar{\pmb{\theta}}_{..}$ is the overall mean that can be computed using $\bar{\pmb{\theta}}_{..}=\frac{1}{2n}\sum_{j=1}^{2n}\bar{\pmb{\theta}}_{.j}$. $s_j^2$ denotes the $j$'th sample sequence variance obtained as $s_j^2=\frac{1}{m-1}\sum_{i=1}^m(\pmb{\theta}_{ij}-\bar{\pmb{\theta}}_{.j})^2$. Utilizing $V_i$ and $B_i$, $R_i$ is defined and computed as \cite{BN2013}:
\begin{equation}
\label{eq:index4}
R_i=\sqrt{\frac{\frac{n-1}{n}V_i+\frac{1}{n}B_i}{V_i}}.
\end{equation}
In theory, the value of $R_i$ equals 1 as $2m \to \infty$. $R_i\gg 1$ indicates that either estimate of the variance can be further decrease by more iterations. In other words, the generated sequences have not yet made a full tour of the target PDF. Alternatively, if $R_i\approx 1$, the sequences are close to the target PDF. Here, following the previous work\cite{BN2013}, a threshold $R_\psi = 1.1$ is adopted to select the value of $M$. Thus, $M\leftarrow 2m$ is set as the number of iterations that satisfy $R_i \leq R_\psi,\forall i$ for the BN.

\section{Numerical Results}\label{result}
\begin{figure}[tbp]
	\centering
	\includegraphics[width=1\linewidth]{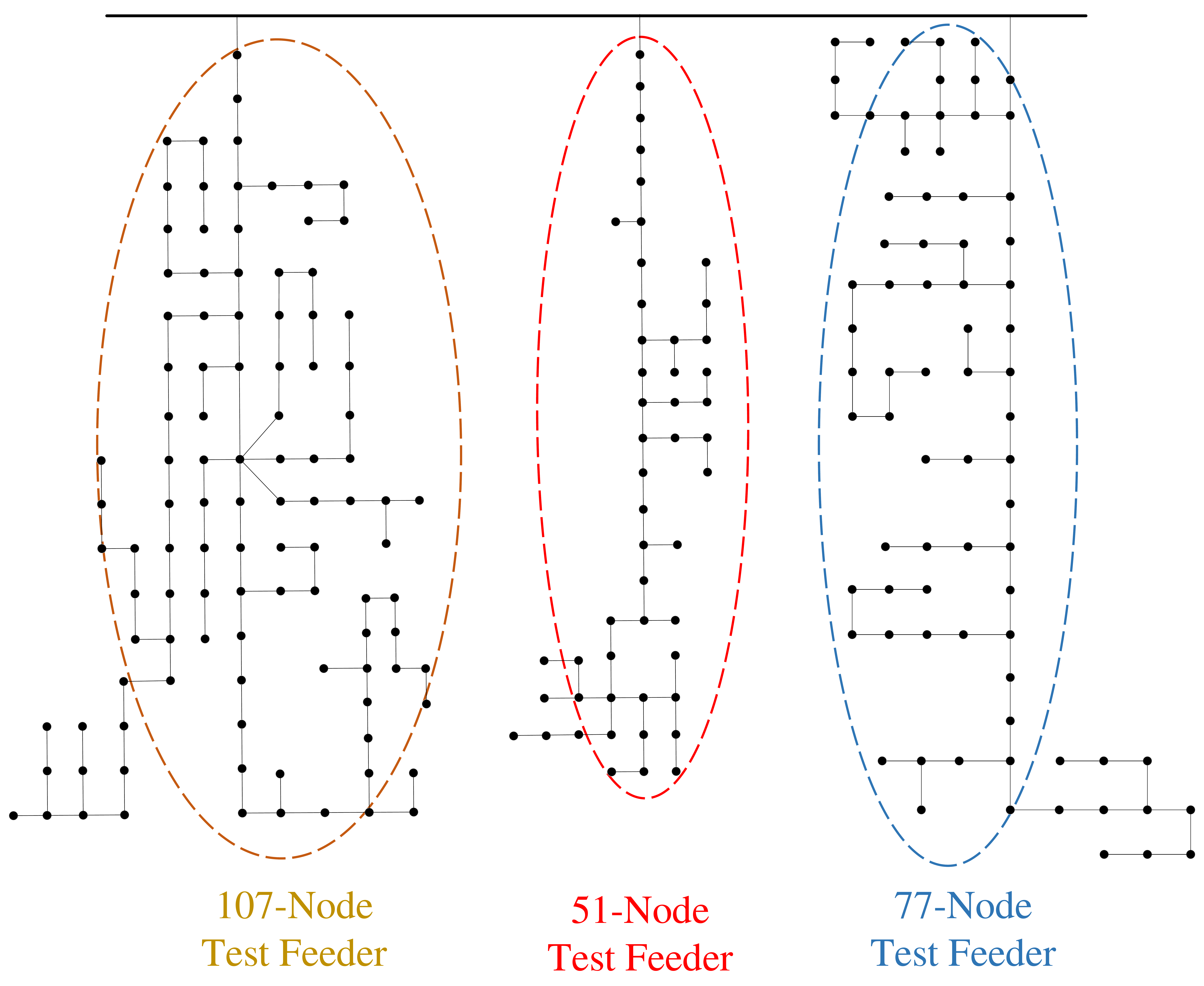}
	\caption{Three test feeders with different sizes.}
	\label{fig:test}
\end{figure}

This section explores the practical effectiveness of the proposed data fusion outage location method. Three real-world distribution feeders are utilized in this case study, which are publicly available online \cite{Test_system}. The topological information is shown in Fig. \ref{fig:test}. All aforementioned evidence, including trouble calls, social media messages, last gasp signal, vegetation information, and wind speed, is utilized in the simulations. Specifically, the human-based evidence is generated using an exponential PDF given $\Delta T$. Note that the parameter of this PDF is considerably different from that of \eqref{eq:ex2} to simulate the uncertainty of the BN parameterization in real-world applications. Consequently, in the outage inference task, we do not know the PDF used to generate evidence and the conditional PDF of the outage location. In this work, the value of $\Delta T$ is assigned as 10 minutes, which indicates that only a fraction of customers are active to give trouble calls or social media messages in our simulations. 
The number of last gasp signals is determined based on the system observability that is calculated as the ratio of customers with SMs to those without SMs. When a customer is assumed to have the SM, this indicates that he is likely to send a last gasp signal when an outage occurs. The vegetation information and the branch's physical parameters are provided by our utility partners. For some unknown parameters, such as tree diameter, we refer to previous work \cite{ouyang2014}. Further, depending on the geographical locations of the available systems, the wind speed data is obtained from national oceanic and atmospheric administration (NOVAA) \cite{noaa}. For each feeder, we have tested the proposed method under three different observability levels, $25\%$, $50\%$, $75\%$. To validate the average performance of the proposed method, a Monte Carlo approach has been utilized to generate $1500$ outage scenarios for each case. In each scenario, a portion of customers are randomly selected to install SMs. Then, the outage location is also randomly chosen. Based on the outage location and the distribution of the SMs, the human- and meter-based evidence is generated.  To simulate real-world power outages, $10\%$, $15\%$, and $3\%$ of total evidence is assumed to be wrong to simulate the illegitimate calls, natural language processing errors, and AMI communication failure.

\begin{figure}[tbp]
	\centering
	\includegraphics[width=0.9\linewidth]{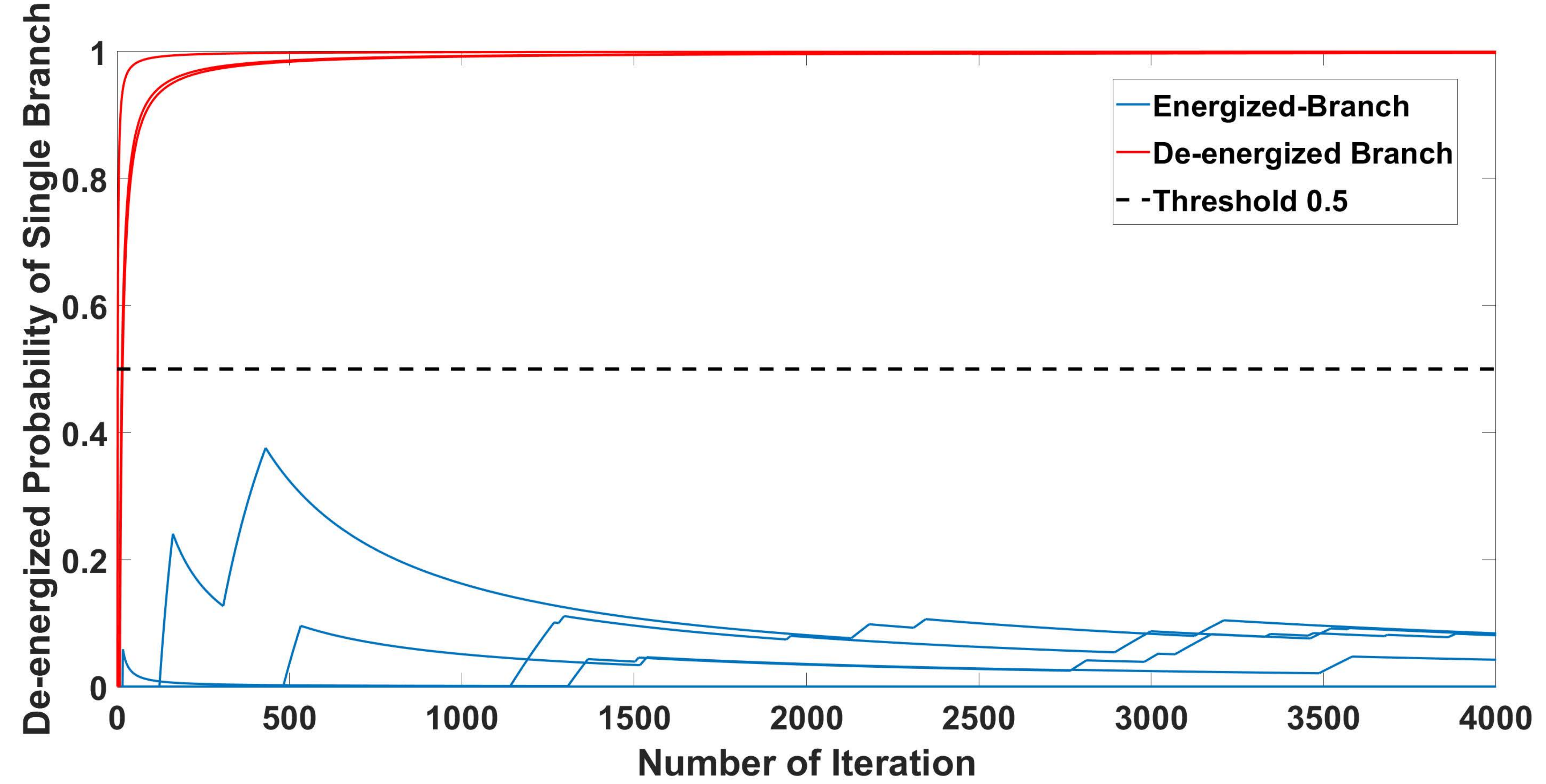}
	\caption{Branch de-energization probabilities for one outage case.}
	\label{fig:example}
\end{figure}

\subsection{Performance of the Proposed Data Fusion Model}\label{conclusion}
Fig. \ref{fig:example} shows the GS-based inferred dis-connectivity probability values of primary branches in the 51-node test feeder in single outage scenario. As can be seen, for branches downstream of the outage location, these probabilities converge to significantly higher values compared to the branches that are not impacted by the outage event. By using the threshold, the energized branches and the de-energized branches can be easily distinguished to locate the outage. This demonstrates that the BN-based outage location inference method is able to correctly determine the state of the system.  

To evaluate the performance of the proposed outage location method for $1500$ generated outage cases in the test systems, several statistical metrics are applied among all branches, including accuracy, precision, recall, and $F_1$ score \cite{sat2006,ZY2017}. These indexes are determined as follows:
\begin{equation}
\label{eq:Accu}
Accuracy=\frac{(TP+TN)}{(TP+FP+FN+TN)}
\end{equation}
\begin{equation}
\label{eq:Prec}
Precision=\frac{(TP)}{(TP+FP)}
\end{equation}
\begin{equation}
\label{eq:recall}
Recall=\frac{(TP)}{(TP+FN)}
\end{equation}
\begin{equation}
\label{eq:f1}
F_1=\frac{(\beta^2+1)*Prec*Recall}{(\beta^2*Prec+Recall)}
\end{equation}
where, TP is the true positive (i.e., state of branch is inferred as de-energized while its actual state is also de-energized), TN is the true negative (i.e., state of branch is considered as an energized while its true state is also energized), FP is the false positive (i.e., state of branch is inferred as de-energized while its actual state is energized), FN is the false negative (i.e., state of branch is inferred as energized while its actual state is de-energized), $P$ and $N$ are the numbers of total positives and negatives, and $\beta$ is the precision weight which is selected to be 1 in this paper. The average values of these indexes are presented in Table. \ref{table:1.1} for the three different test feeders with various observability levels. In all cases, the lowest accuracy, precision, recall, and $F_1$ score are $98.7\%$, $83.47\%$, $98.88\%$, and $88.05\%$, respectively. For $50\%$ and $75\%$ observability cases, all branch-level indexes reach values over 0.9. Also, the system-level accuracy is calculated as the percentage of times that the states of all the branches/customers have been inferred correctly in outage scenarios. In other words, even though the outage location is inferred correctly, the system-level index may fail because of one misclassified branch. As shown in the table, when the observability is 25$\%$, the system-level accuracy is about 70$\%$. This could be due to the evidence scarcity. We have analyzed the failed scenarios. In more than 80$\%$ of these scenarios, the proposed method can infer the actual location of the outage but misjudged the status of one or two branches. For the cases that have $75\%$ observability, the system-level accuracy is about 90$\%$. This result is not surprising since we have assigned false positive and false negative alarms in each scenario. Such alarms reduce the completeness of outage information. By comparing the results of the three feeders, it can be concluded that the performance of the proposed outage location method improves as the observability increases, due to the high confidence levels of meter-based evidence. Also, the proposed algorithm shows almost the same level of performance over the different test feeders. This result demonstrates that the BN-based outage location method is nearly insensitive to the topology of the underlying network. 

\begin{figure}[tbp]
	\centering
	\includegraphics[width=3.5in]{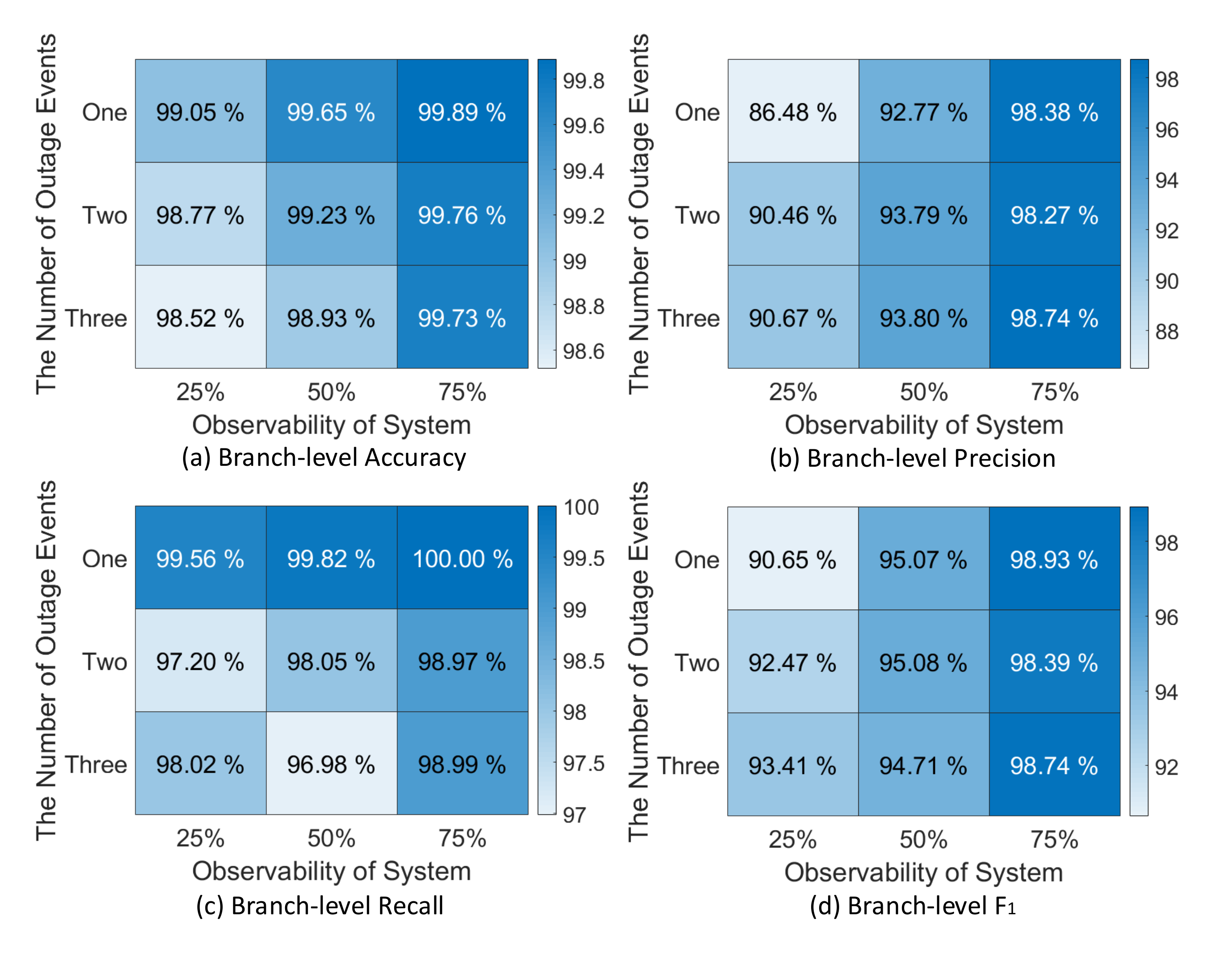}
	\caption{Sensitivity analysis with coinciding multi-outage events.}
	\label{fig:multi}
\end{figure}

To further evaluate the performance of our method, coinciding multiple outage events are generated in three test systems. The locations of the outages are randomly selected. For concreteness, we have also calculated the accuracy under $25\%$, $50\%$, and $75\%$ observability levels. Fig. \ref{fig:multi} shows the performance indexes as a function of observability level and the number of outages. As can be seen, almost in all cases, higher observability improves the performance indexes regardless of the number of coinciding outage events. Also, the indexes have nearly similar values in cases with single and multiple outages. Hence, we can conclude that the method has a stable performance for multiple outages.

\subsection{GS Calibration Results}\label{conclusion}
\begin{figure}[tbp]
	\centering
	\includegraphics[width=0.9\linewidth]{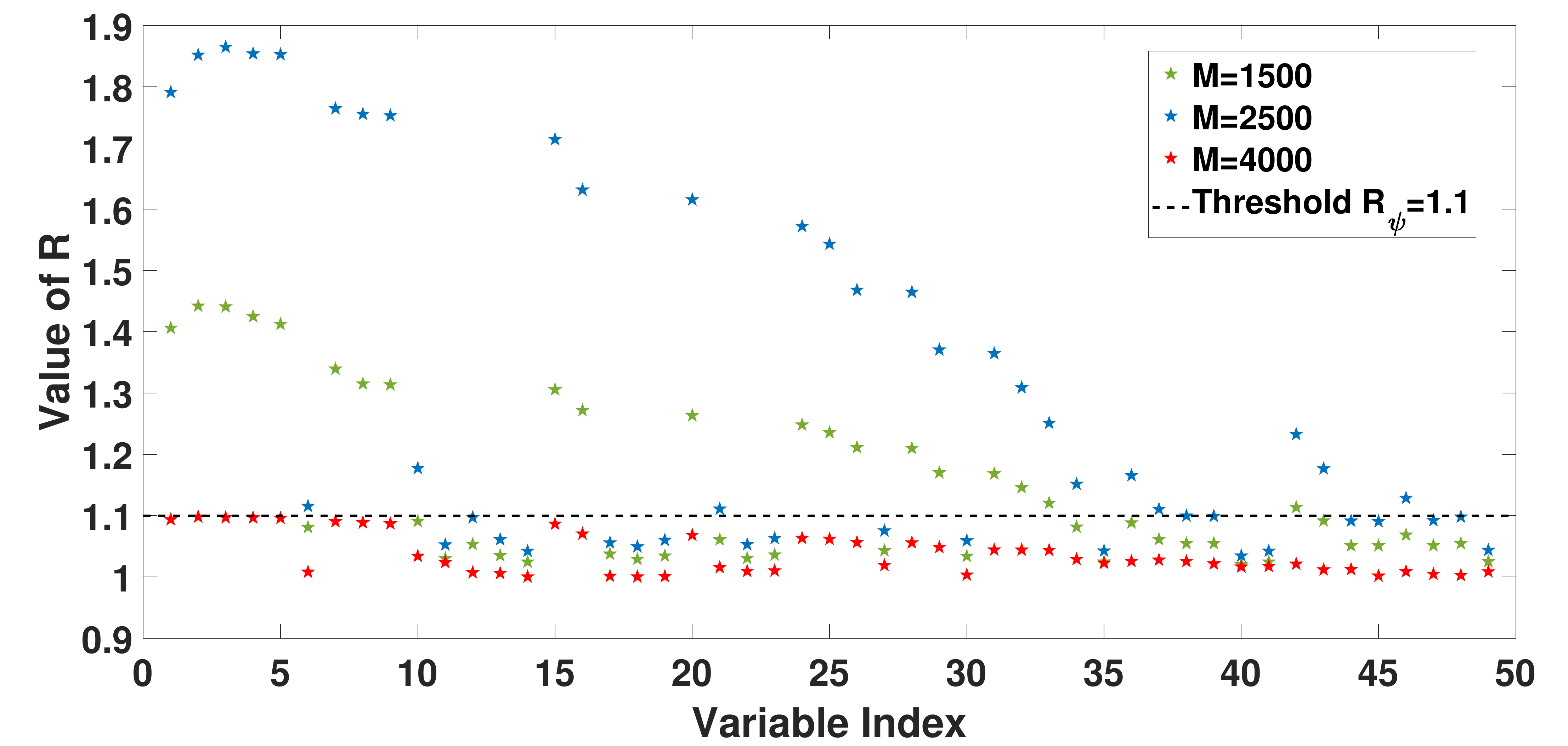}
	\caption{GS algorithm calibration results for the 51-node system.}
	\label{fig:cd}
\end{figure}

Basically, the GS calibration is a trial and error process using a specific index, $R$. Hence, in each test feeder, we have generated 500 sample sequences for each unknown variable in the BN at different sampling iterations, $M$. Fig. \ref{fig:cd} shows the values of $R_i$ in the $51$-node test feeder. As can be seen, by increasing the number of $M$, the values of $R_i$'s tend to converge to 1. By selecting $M=4000$, all $R_i$'s drop below the user-defined calibration threshold, $R_\psi = 1.1$, which indicates that GS has reached a reasonable number of iterations in this BN. Note that GS calibration is a offline process; as a result, the high computational burden of the trial and error process does not impact the real-time performance of the proposed method.

\begin{figure}[tbp]
	\centering
	\includegraphics[width=0.9\linewidth]{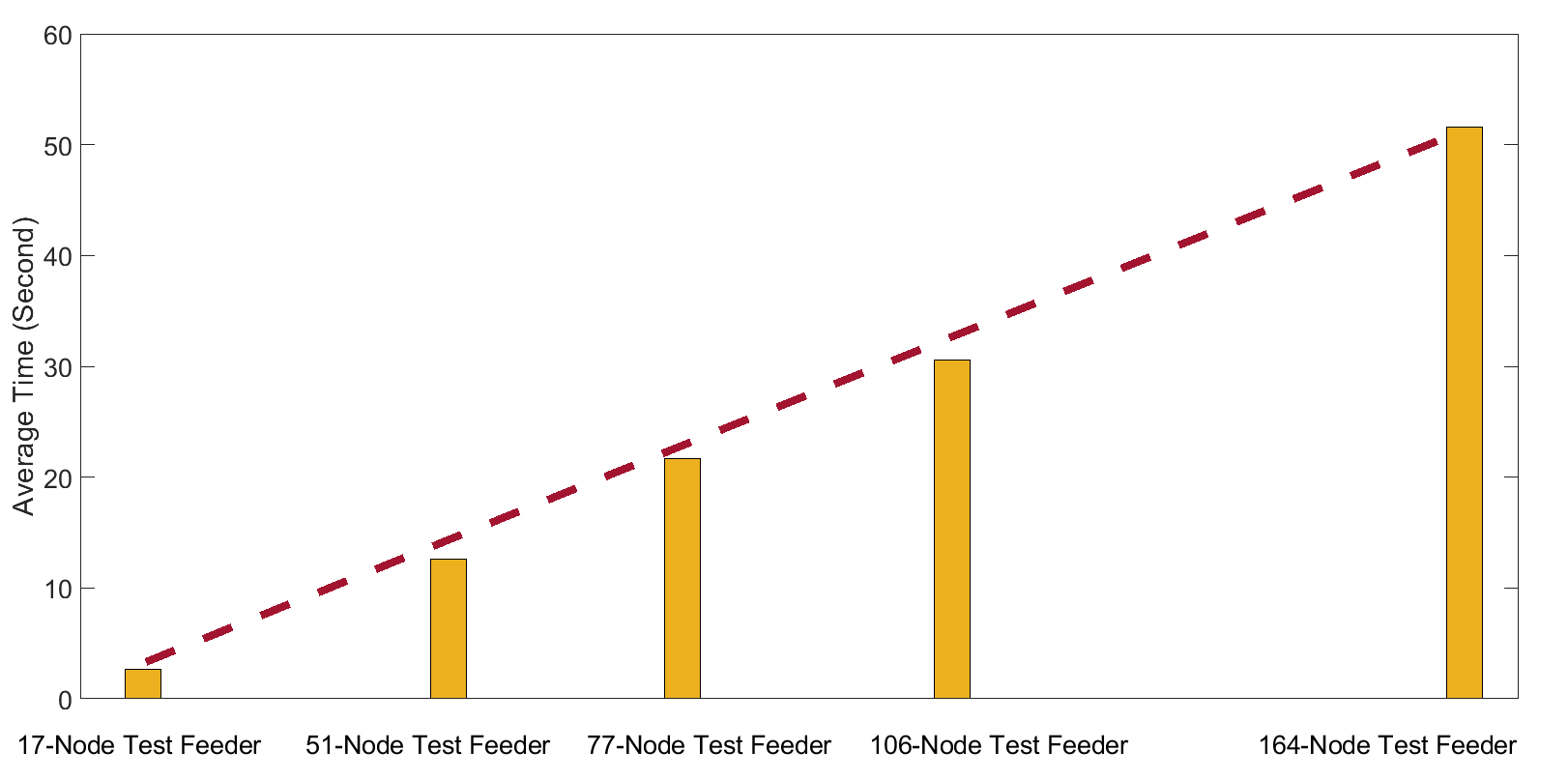}
	\caption{Average simulation time for the three test feeders.}
	\label{fig:time}
\end{figure}

\subsection{Computational Complexity Analysis}
The case study is conducted on a standard PC with an Intel(R) Xeon(R) CPU running at 4.10GHZ and with 64.0GB of RAM and an Nvidia Geforce GTX 1080ti 11.0GB GPU. To provide a comprehensive computational complexity analysis, the proposed method is conducted on two additional real-world distribution feeders: a 17-node and 164-node feeders. The detailed information of these feeders can be found in \cite{yuanGAN} and \cite{yyx2019}. Fig. \ref{fig:time} shows the average computational time of outage inference for the test feeders. As described in the figure, by using our standard PC, the average computational time for outage location inference in five test feeders are $\{2.7s,12.58s,21.64s,30.14s,51.59s\}$, respectively. Further, due to significant complexity reduction by the proposed BN, it can be observed that the computational complexity and the size of the distribution feeders have almost a linear relationship rather than an exponential one. Also, the proposed model does not infer outage location in a system-wide fashion, but performs feeder-level location estimation. This strategy enables parallel computation of different feeders to further reduce the computational time. These salient features can facilitate the application of practical distribution systems.

\section{Conclusion}\label{conclusion}
In this paper, we have presented a novel multi-source data fusion approach to detect and locate outages in partially observable distribution networks. The problem is cast as the process of inferring the probabilities of post-event operational topology candidates. Our method encodes the network's topology and the causal relationship between outage evidence and branch states into BNs by leveraging the conditional independence inherent in distribution grids. By constructing the BNs, the proposed method is able to infer the connectivity probability of individual primary branches with nearly linear complexity in the size of the network. Moreover, this method exploits data redundancy to reduce the impact of data uncertainty, and is suitable for arbitrary radial distribution systems. Based on simulation results on real-world networks, the proposed method can accurately detect and locate outage events within a short time.

Future study will seek to extend the proposed method in meshed grids with distributed energy resources. BNs alone cannot fully capture conditional independencies when there are multi-directional power flows. Hence, we plan to explore hybrid graphs that consist of both directed BNs and fully undirected Markov networks. Further, a joint Boltzmann distribution function will be investigated to embody graph parameters. 

\ifCLASSOPTIONcaptionsoff
  \newpage
\fi



\bibliographystyle{IEEEtran}
\bibliography{IEEEabrv,./bibtex/bib/IEEEexample}
\end{document}